\documentclass[%
reprint,
 amsmath,amssymb,
 aps,
 pre,
floatfix
]{revtex4-1}

\usepackage{graphicx}
\usepackage{dcolumn}
\usepackage{bm}
\usepackage{url}
\usepackage{hyperref}
\usepackage{pslatex} 

\newcommand{\vect}[1]{\boldsymbol{#1}}

\begin{document}

\preprint{APS/123-QED}
\title{Constrained information flows in temporal networks reveal intermittent communities}

\author{Ulf Aslak}
    \email{ulfaslak@gmail.com}
    \affiliation{
        Centre for Social Data Science, University of Copenhagen, DK-1353 K\o{}benhavn K
    }
    \affiliation{
        DTU Compute, Technical University of Denmark, DK-2800 Kgs.~Lyngby
    }
\author{Martin Rosvall}
    \email{martin@email.edu}
    \affiliation{
        Integrated Science Lab, Department of Physics, Ume{\aa} University, SE-901 87 Ume{\aa}, Sweden
    }
\author{Sune Lehmann}
    \email{sljo@dtu.dk}
    \affiliation{
        DTU Compute, Technical University of Denmark, DK-2800 Kgs.~Lyngby
    }
    \affiliation{
        Niels Bohr Institute, University of Copenhagen, DK-2100 K\o{}benhavn \O
    }
    \affiliation{
        Department of Sociology, University of Copenhagen, DK-1353 K\o{}benhavn K
    }

\date{\today}

\begin{abstract}
Many real-world networks represent dynamic systems with interactions that change over time, often in uncoordinated ways and at irregular intervals. 
For example, university students connect in intermittent groups that repeatedly form and dissolve based on multiple factors, including their lectures, interests, and friends. 
Such dynamic systems can be represented as multilayer networks where each layer represents a snapshot of the temporal network.
In this representation, it is crucial that the links between layers accurately capture real dependencies between those layers.
Often, however, these dependencies are unknown. 
Therefore, current methods connect layers based on simplistic assumptions that do not capture node-level layer dependencies. 
For example, connecting every node to itself in other layers with the same weight can wipe out essential dependencies between intermittent groups, making it difficult or even impossible to identify them.  
In this paper, we present a principled approach to estimating node-level layer dependencies based on the network structure within each layer. 
We implement our node-level coupling method in the community detection framework Infomap and demonstrate its performance compared to current methods on synthetic and real temporal networks. 
We show that our approach more effectively constrains information inside multilayer communities so that Infomap can better recover planted groups in multilayer benchmark networks that represent multiple modes with different groups and better identify intermittent communities in real temporal contact networks. 
These results suggest that node-level layer coupling can improve the modeling of information spreading in temporal networks and better capture intermittent community structure.
\end{abstract}
\maketitle

Temporal network representations of dynamic complex systems allow researchers to describe changing interaction patterns.
Increasingly, high-resolution interaction data require methods that can simplify and highlight important temporal network structures.
An important category of such structures is highly intraconnected groups of nodes, so-called communities.
If the nodes represent individuals who alternate between various roles in social temporal networks, the communities will repeatedly form and dissolve at multiple temporal scales in an intermittent way.
A simple approach to identify intermittent communities is to first separate a temporal network into a sequence of static snapshots, that is, a multilayer network~\cite{kivela2014multilayer, de2013mathematical}, then independently cluster each layer, and finally match the communities across the layers to find the temporal communities~\cite{palla2007quantifying, tantipathananandh2007framework, pietilanen2012dissemination, kauffman2014dyconet, he2015fast, sekara2016fundamental}. 
Other approaches, including three-way matrix factorization~\cite{gauvin2014detecting}, time-node graphs~\cite{speidel2015community}, and stochastic block models~\cite{peixoto2015inferring, gauvin2014detecting, matias2016statistical}, can directly cluster the multilayer network, but are also unable to incorporate explicit dependencies between layers. 
To take into account such interdependencies, some methods cluster multilayer networks using interlayer links that represent specific causal or correlational dependencies between the layers~\cite{mucha2010community, chen2013detecting, de2015identifying, bazzi2016community}.
However, explicit interlayer dependencies are often not available to researchers. 
Moreover, current approaches for estimating such dependencies by, for example, comparing independently inferred community structure between layers~\cite{larremore2013network}, using stochastic block modeling~\cite{stanley2016clustering}, or applying link prediction through cross-validation~\cite{de2017community}, consider only dependencies  between entire layers.
In contrast, real systems with multiple and asynchronous recurrent events generate dependencies between layers with varying strength within layers. 
By ignoring such node-level dependencies, current methods wash out important dependencies in multilayer networks with intermittent communities at multiple temporal scales.

In this paper, we present a flow-based method that first couples node pairs in different layers based on the similarity between their network neighborhood flow patterns, and then -- based on the network structure within layers combined with these node-level interlayer dependencies -- identifies temporal communities in the resulting multilayer network. 
For a single node, non-overlapping neighborhoods are not coupled and identical neighborhoods are maximally coupled.
In a social network, this neighborhood flow coupling captures that individuals typically share similar information in similar social contexts.
In this sense, neighborhood flow coupling models causal dependencies across time. 
Finally, we adapt the flow-based community detection algorithm Infomap~\cite{rosvall2008maps, rosvall2009map} to make use of this information.
We demonstrate the usefulness of neighborhood flow coupling for multilayer community detection on benchmark networks. Additionally, we reveal and visualize the temporal evolution of intermittent communities in two temporal human contact networks~\cite{genois2015data, stopczynski2014measuring}.
While our method targets intermittent communities in temporal contact networks represented by multilayer networks, it nevertheless outperforms other methods in standard benchmark tests on multilayer networks.

\section*{Methods}
\begin{figure}
    \centering
    \includegraphics[width=\linewidth]{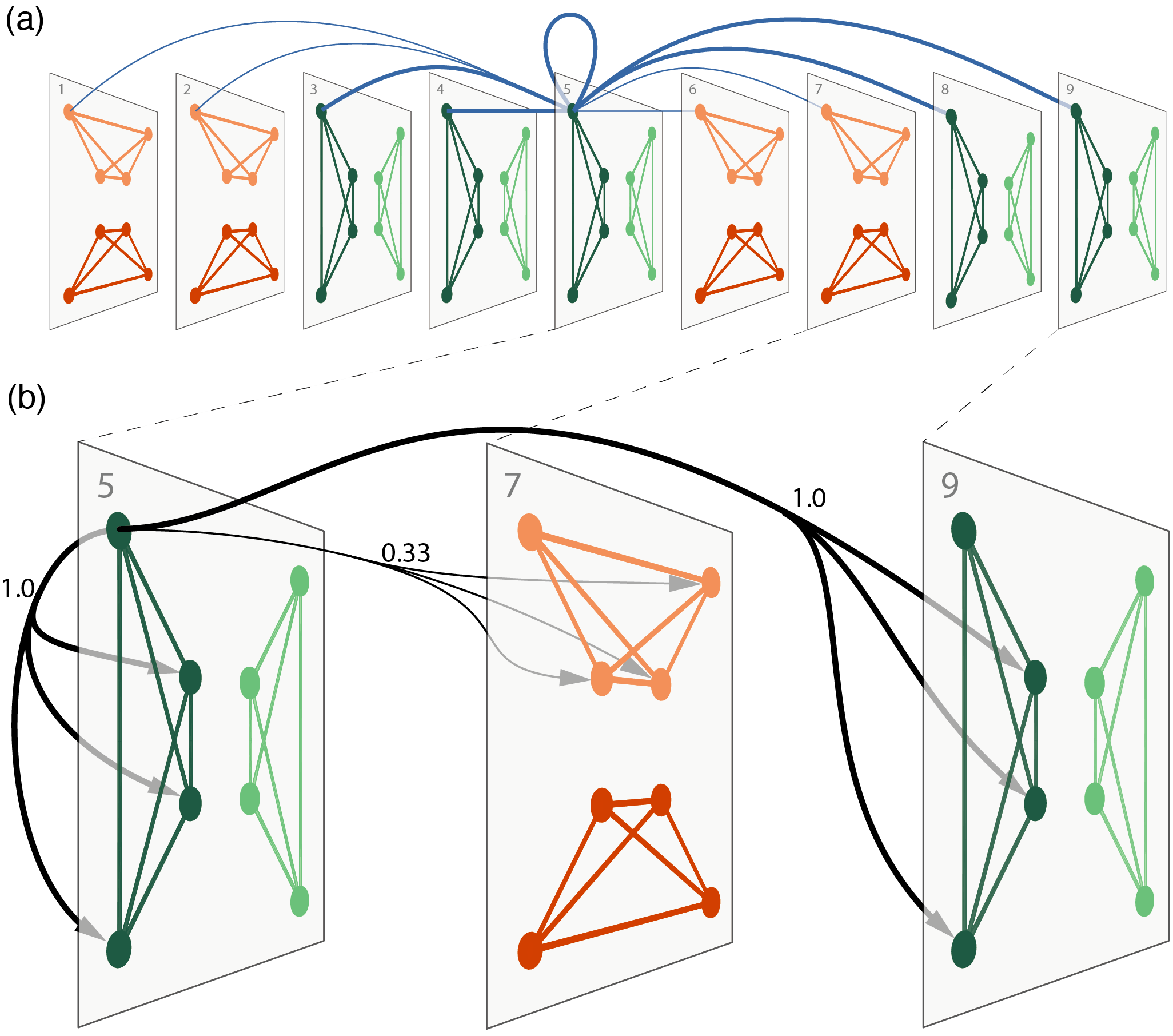}
    \caption{\label{fig:figures_two_step_approach}\textbf{Neighborhood flow coupling between a state node and its sibling states in a multilayer network}. (a) Interlayer coupling, $D_i^{\alpha\beta}$, from the top left state node in layer 5 to all state nodes of the same physical node. State nodes with more similar intralayer outlinks couple more strongly, indicated by the stroke width. (b) Interlayer links, $P_{ij}^{\alpha\beta}$, are directed and connect a state node to neighbors of state nodes of the same physical node, with weight proportional to the coupling strength and the intralayer link weight of the neighbor node (Eq.~\ref{eq:interlayer_link_weight}).}
\end{figure}

In complex networks, groups of nodes in which flows stay for a long time provide a useful notion of communities~\cite{rosvall2008maps,delvenne2010stability}.
Such communities also can provide straightforward generalizations to multilayer networks~\cite{de2015identifying}. 
We use multilayer networks with \textit{physical} nodes and \textit{state} nodes.
Physical nodes represent system components, while state nodes, one for each physical node and layer, represent constraints on flows (see Fig.~\ref{fig:figures_two_step_approach}).  
Accordingly, we consider multilayer communities to be groups of state nodes that capture flows for a significantly long time. 
In this way, assigning a physical node's state nodes to different communities naturally results in overlapping communities.

In real-world temporal networks, communities often form and dissolve multiple times with a shorter presence than absence~\cite{sekara2016fundamental}.
From the perspective of the entire network, these intermittent communities are often asynchronous in the sense that each community forms and dissolves independently in time relative to other communities. 
Examples of intermittent communities include group voting trends in the US Senate~\cite{mucha2010community}, time-dependent sets of correlated financial assets~\cite{bazzi2016community}, social cores in contact networks~\cite{sekara2016fundamental}, and modules of coherently active brain areas~\cite{braun2015dynamic}. 
Because nodes in these intermittent communities are not able to share information across their long absent times -- since nodes are unlikely to be connected -- current methods for identifying communities with long flow-persistence times cannot effectively capture the potential for information transfer.
A causal dependency across time requires that interlayer link strengths represent the degree to which information is likely to flow between state nodes in adjacent as well as distant layers. 
Some existing methods indeed evaluate dependencies between layers, but they do it by coupling entire layers~\cite{larremore2013network, stanley2016clustering, de2017community}. 
There are two important drawbacks to this approach. 
First, coupling a physical node's state nodes across all layers generates a large number of links, resulting in computational challenges. 
For example, in a network with $n$ nodes in $t$ layers with average degree $\left<k\right>$, we need $\left<k\right>t^2n$ links in addition to the within-layer links in order to represent connections between state nodes.
Second, for large networks with many time slices and intermittent and asynchronous communities, the uniform interlayer links can also dilute community boundaries and aggregate distinct communities (we will discuss this point in detail below).
To counter these drawbacks of uniform linking, we propose interlayer dependencies at the node level.
By forming state-node-specific interlayer links, neighborhood flow coupling generates high-resolution yet sparse multilayer networks that can capture intermittent communities.

\subsection*{Neighborhood flow coupling}
The goal of our flow-based approach is to enable interlayer coupling based on the local structural properties of the multilayer network. 
Each layer's intralayer link structure represents the constraints on network flows at a given time or state of the system. 
Specifically, we model the network flows in each layer with a random walker that moves from state node to state node guided by the outgoing intralayer links.
Because the links represent where flows can move, similar outgoing intralayer link flows in two state nodes of a physical node suggests that the state nodes represent similar states of the physical node. 
In a social setting, for example, the same group of people may meet again and take up where they left off last time they met. 
More precisely, the more similar the within-layer flow patterns are, the less would the constraints change and the less would information be lost if the two state nodes were lumped together.
We use this information loss measure to couple layers: The less information that is lost if the state nodes were combined, the stronger the interlayer coupling between the state nodes. 

This neighborhood flow coupling based on the information loss from merging state nodes is accurately captured by the Jensen-Shannon divergence. 
In detail, for neighborhood flow coupling between physical node $i$'s state nodes in layers $\alpha$ and $\beta$, the state nodes' normalized intralayer outlinks $\vect{P}_i^\alpha$ and $\vect{P}_i^{\beta}$ give their coupling strength $D_i^{\alpha\beta}$,
\begin{align}
\label{eq:interlayer_link_weight}
D_i^{\alpha\beta} &= 1 - \mathrm{JSD}\left(\vect{P}_i^\alpha, \vect{P}_i^{\beta}\right)\\
&= 1 - H\left(\frac{1}{2}\vect{P}_i^\alpha + \frac{1}{2}\vect{P}_i^{\beta}\right) + \frac{1}{2}H\left(\vect{P}_i^\alpha\right) + \frac{1}{2}H\left(\vect{P}_i^\beta\right),
\end{align}
where $\mathrm{JSD}(\cdot,\cdot)$ is the Jensen-Shannon divergence and $H(\cdot)$ is the Shannon entropy. 
In a multilayer network with neighborhood flow coupling, a random walker moves from state node to state node within a layer guided by the intralayer links and, at rate $r$, transitions to any layer, including the currently visited layer, proportional to the intralayer link similarity between the state nodes (see Fig.~\ref{fig:figures_two_step_approach}a). We include interlayer links to the same layer because they allow for generalizations with complete layer information at rate $1-r$ and no layer information when the layer constraints are relaxed at rate $r$, as if the layers were aggregated.

Neighborhood flow coupling disregards the temporal ordering of layers. However, for longer time-scales or depending on the research question at hand, layer coupling that depends on temporal distance can be implemented. For example, Eq.\ \eqref{eq:interlayer_link_weight} can be scaled by a factor that depends on the temporal distance between layers.

In any case, intralayer \textit{links} connect state nodes to their neighbors within the same layer and interlayer \textit{coupling} connects state nodes of the same physical node in different layers. 
For example, take a random walker at a state node of physical node $i$ in layer $\alpha$, $(i,\alpha)$ for short.
With probability $1-r$ it remains in the same layer and moves to state node $(j,\alpha)$ with probability proportional to the intralayer link weight $W_{ij}^\alpha$.
With the remaining probability $r$ it relaxes the layer constraint, switches to any layer $\beta$ proportional to the interlayer coupling strength $D_i^{\alpha\beta}$, and moves to state node $(j,\beta)$ proportional to the intralayer link weight $W_{ij}^\beta$. 
Consequently, with intralayer out-strength $s_i^\beta = \sum_j{W_{ij}^\beta}$ and interlayer out-strength $S_i^\alpha = \sum_\beta{D_i^{\alpha\beta}}$ of state node $(i,\alpha)$, the transition probabilities as a function of $r$ are
\begin{align}
    \label{eq:transition_probability}
    P_{ij}^{\alpha\beta}(r) = (1-r) \frac{W_{ij}^\beta}{s_i^\beta} \delta_{\alpha\beta} + r \frac{D_i^{\alpha\beta}}{S_i^\alpha} \frac{W_{ij}^\beta}{s_i^\beta},
\end{align}
where $\delta_{\alpha\beta}$ is the Kronecker delta. Therefore, relaxing the layer constraint means that the random walker loses memory of which layer it is currently visiting and instead follows the outgoing links of any state node of the same physical node. 
With uniform interlayer coupling, relaxing the layer constraint corresponds to a step on the fully aggregated network. 
However, neighborhood flow coupling takes advantage of higher-order information in the multilayer network that enables longer persistence times in intermittent communities (see Fig.~\ref{fig:figures_two_step_approach}b).

\subsection*{Neighborhood flow coupling and the map equation}
To use neighborhood flow coupling in the context of community detection, we use the map equation framework for multilayer networks~\cite{rosvall2009map, de2015identifying}.
For our purposes, the map equation framework comes with two advantages. 
Firstly, the map equation is flow-based and directly integrates state-node-specific interlayer flows, as it balances intralayer and interlayer flows by relaxing the intralayer constraints with an interlayer relax rate. 
Secondly, the map equation naturally clusters coupled state nodes with similar intralayer links in the same community, as it assigns state nodes of the same physical node and community to the same codeword to capture the fact that they represent the same physical object.
Therefore, the flow-based and information-theoretic nature of the map equation is a good fit with neighborhood flow coupling.

In detail, for a two-level modular description of flows from node to node in $m$ communities, one \emph{index codebook} contains the community-enter codewords and $m$ \emph{module codebooks} contain the node-visit and community-exit codewords within modules. 
Each codebook's average codeword length is given by the Shannon entropy of their rates of use, $\mathcal{Q}$ for enter codewords with total rate of use $q_\curvearrowleft$, and $\mathcal{P}_j$ for codewords in community $j$ with total rate of use $p_{j\circlearrowright}$. 
For node partition $\mathsf{M}$, the map equation therefore takes the form
\begin{align}
    \label{eq:themapequation}
    L\left(\mathsf{M}\right) = q_\curvearrowleft H(\mathcal{Q}) + \sum_{j=1}^m p_{j\circlearrowright}H(\mathcal{P}_j).
\end{align}
Applied to a possibly weighted and directed network, Infomap searches for the node partition $\mathsf{M}$ that minimizes the map equation and reveals the most modular regularities in the network flows. 

The map equation remains the same for multilayer networks, with one important generalization: when state nodes of the same physical node are assigned to the same community, they are assigned a common code word derived from their total visit rate. 
This coding scheme captures the very essence of multilayer networks, that all state nodes of the same physical node represent the same physical object~\cite{de2015identifying}. 

We have implemented the neighborhood flow coupling in the Infomap software package available on \url{www.mapequation.org}. Neighborhood flow coupling is activated with the flag \texttt{--multilayer-js-relax-rate}. For memory efficiency or for encoding temporal ordering of layers, interlayer links can be thresholded based on the Jensen-Shannon divergence and temporal distance between layers with the flags \texttt{--multilayer-js-relax-limit} and \texttt{--multilayer-relax-limit}, respectively.

Neighborhood flow coupling can be used with other community detection frameworks such as multilayer modularity optimization~\cite{mucha2010community}. While multilayer modularity cannot distinguish state nodes from physical nodes,  high density of interlayer links between similar layers will nevertheless make it easier to identify intermittent communities.  Moreover, the basic principle of neighborhood flow coupling extends beyond community detection and can be useful for capturing spreading processes in multilayer networks when interlayer coupling information is absent.

\section*{Results}
We first validate the performance of Infomap with neighborhood flow coupling on benchmark networks with multilayer structure. 
Then we identify temporal communities in two face-to-face contact networks.

\subsection*{Performance tests on benchmark networks} 
We compare neighborhood flow coupling with other interlayer coupling schemes on three types of multilayer benchmark networks to test each method's ability to handle overlapping community structure, recover intermittent communities in increasingly sparse multilayer networks, and retain flows within intermittent communities. 
We compare neighborhood flow coupling (NFC) with full coupling (FC), adjacent coupling (AC), and no-coupling (NC). 
Full coupling with uniform coupling across layers and no-coupling with only the intrinsic coupling from the multilayer coding scheme are extreme cases of neighborhood flow coupling, when the structural similarity in Eq.~\eqref{eq:interlayer_link_weight} is either 1 or 0 across all state nodes of the same physical node~\cite{de2015identifying}. 
Adjacent coupling with uniform coupling strength to the nearest layers is an appealing method for gradually changing communities, but cannot capture intermittent communities. 
These alternative coupling methods provide references to compare and contrast the results of neighborhood flow coupling.

\subsubsection*{Community overlap}
\begin{figure}
    \centering
    \includegraphics[width=\linewidth]{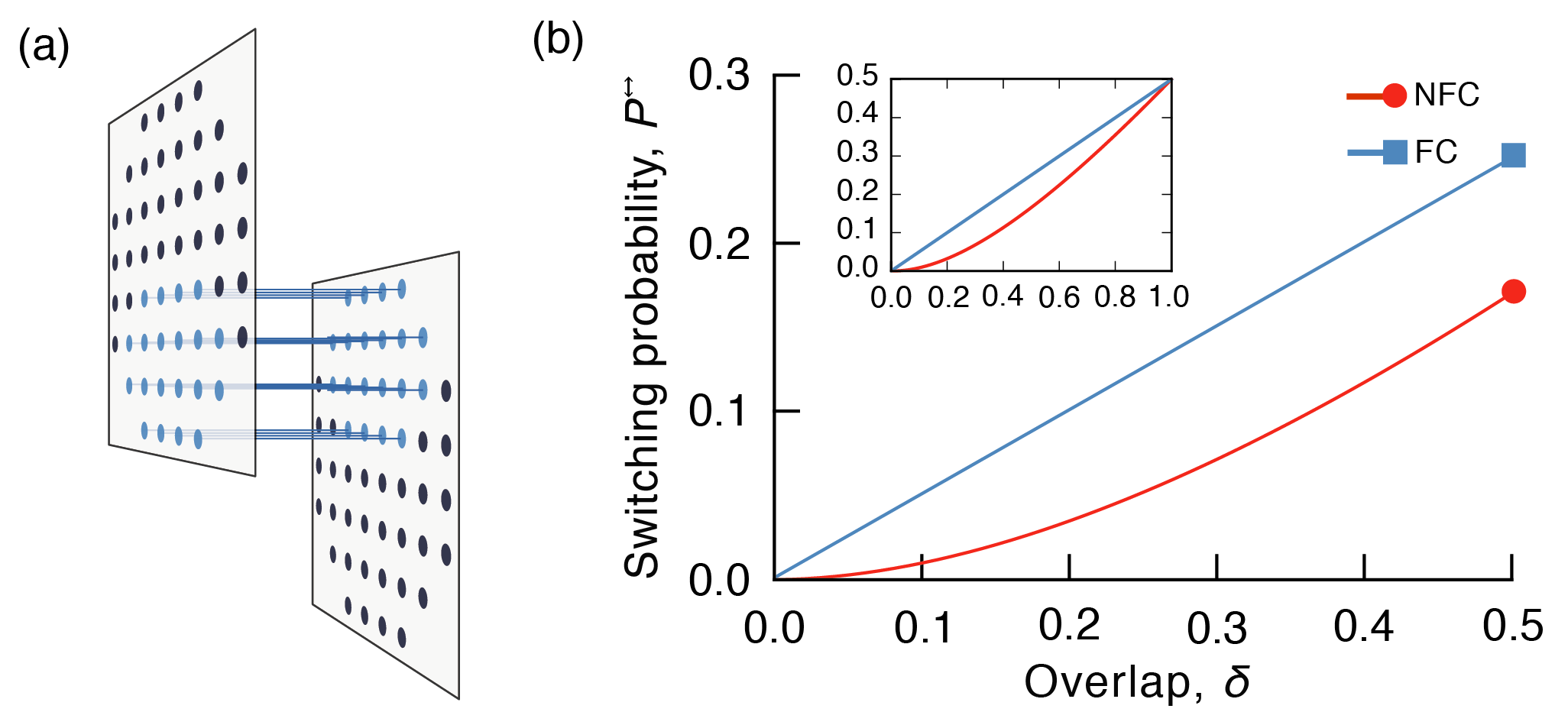}
    \caption{\label{fig:simulated_overlap_conceptual} \textbf{Coupling strength as a function of overlap}. (a) A conceptual illustration of two identical fully connected communities of size $N=52$ that reside in adjacent layers and overlap by $\delta N=20$ nodes. Overlapping nodes are in blue, and self-couplings are omitted for the purpose of illustration. (b) Analytical layer switching probability for full and neighborhood flow coupling as functions of overlap computed for $r=1$ and $N\gg1$. The inset shows the full range from 0 to 1.}
\end{figure}

In real networks, such as face-to-face networks, communities are rarely completely non-overlapping but instead will share some members. 
Therefore, we investigate how neighborhood flow coupling handles overlap compared to full coupling. 
We begin by considering the simplest possible example: two identical, fully connected communities of size $N$ that overlap by a fraction $\delta$ (Fig.~\ref{fig:simulated_overlap_conceptual}a). 
In this network, a random walker traversing the network occupies a node $i$ inside the overlap with probability $\delta$ and performs a relax step with probability $r$. 
Consequently, the random walker switches layer with probability
\begin{align}
    \label{eq:layer_switch}
    P^\leftrightarrow=\delta r \dfrac{D_i^{\alpha\beta}}{D_i^{\alpha\beta} + 1}.
\end{align}
A higher $P^\leftrightarrow$ corresponds to stronger coupling between the two communities and increased preference for classifying the two as a single community. 
For full or adjacent coupling, $D_i^{\alpha\beta} = 1$ and $P^\leftrightarrow_{FC}=1/2 \delta r$.
For neighborhood flow coupling, $D_i^{\alpha\beta}$ from \eqref{eq:interlayer_link_weight} for a node in the overlap yields
\begin{align}
    \label{eq:layer_switch_NFC}
    P^\leftrightarrow_{NFC} = \delta r \dfrac{(\delta N - 1)}{(\delta N - 1) + (N - 1)}.
\end{align}
That is, the probability of switching layers is the fraction of time steps that a random walker can switch, $\delta r$, multiplied by the probability that a relax step will result in a layer switch (which is the number of nodes the walker can reach in the other layer divided by the total number of nodes the walker can reach when inside the overlap). Note that when $N \gg 1$ the probability of switching layers is only a function of $\delta$ and $r$.

\begin{figure}
    \centering
    \includegraphics[width=\linewidth]{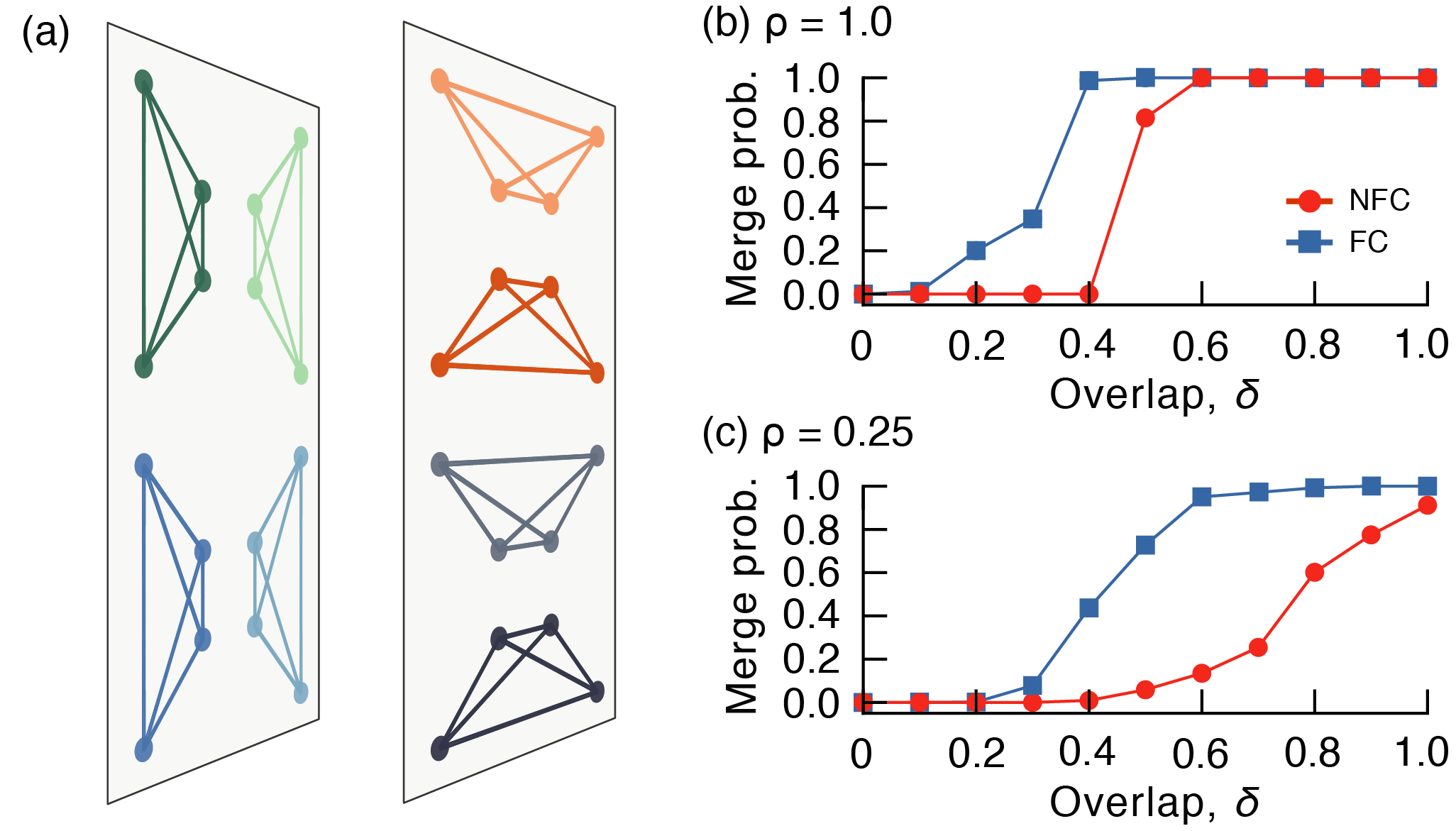}
    \caption{\label{fig:simulated_overlap} \textbf{Full coupling merges communities at lower overlap}. (a) Schematic illustration of two layers of cliques with variable overlap. (b) The probability that Infomap with different coupling schemes merges two communities in separate layers as a function of their node overlap. (c) Same as (b) but for sparser networks, in which all but a fraction of $\rho=0.25$ links are randomly removed.}
\end{figure}

Figure~\ref{fig:simulated_overlap_conceptual}b shows $P^\leftrightarrow$ as a function of $\delta$ for full and neighborhood flow coupling at relax rate $r=1$ and $N\gg 1$. 
This test shows that the layer switching probability can differ significantly in the important range $\delta \in ]0, 0.5]$, suggesting that neighborhood flow coupling has a lower tendency to merge overlapping communities compared to both full coupling and adjacent layer coupling.

To compare neighborhood flow coupling and full coupling in a more complex scenario, we measure the threshold of overlap at which the two methods collapse overlapping communities. First, we construct a two-layer network benchmark model with 500 physical nodes partitioned into 50 communities of uniform size 10, where the communities in each layer differ by some number of random edge swaps (Fig.~\ref{fig:simulated_overlap}a). 
Then, using Infomap with both coupling schemes on 1000 network realizations, we record the overlap for each pair of communities in different layers and whether or not Infomap merges them (Fig.~ \ref{fig:simulated_overlap}b).
We see that full coupling merges communities more aggressively than neighborhood flow coupling, in some cases even when they only overlap by two nodes. Conversely, Infomap with neighborhood flow coupling requires substantial overlap before it merges two distinct communities. 

In real-world networks, communities are sometimes sparsely linked internally.
Since the neighborhood flow coupling considers overlap in internal link structure rather than in nodes, partly overlapping communities will merge with lower probability when the communities are sparser. 
For example, with all but a fraction $\rho = 0.25$ of the edges randomly removed from each community, the merge probability decreases more for neighborhood flow coupling than full coupling (Fig. \ref{fig:simulated_overlap}c). 
In networks with few layers, the network under study and the research question at hand should determine which coupling method is best. 

\begin{figure}
    \centering
    \includegraphics[width=\linewidth]{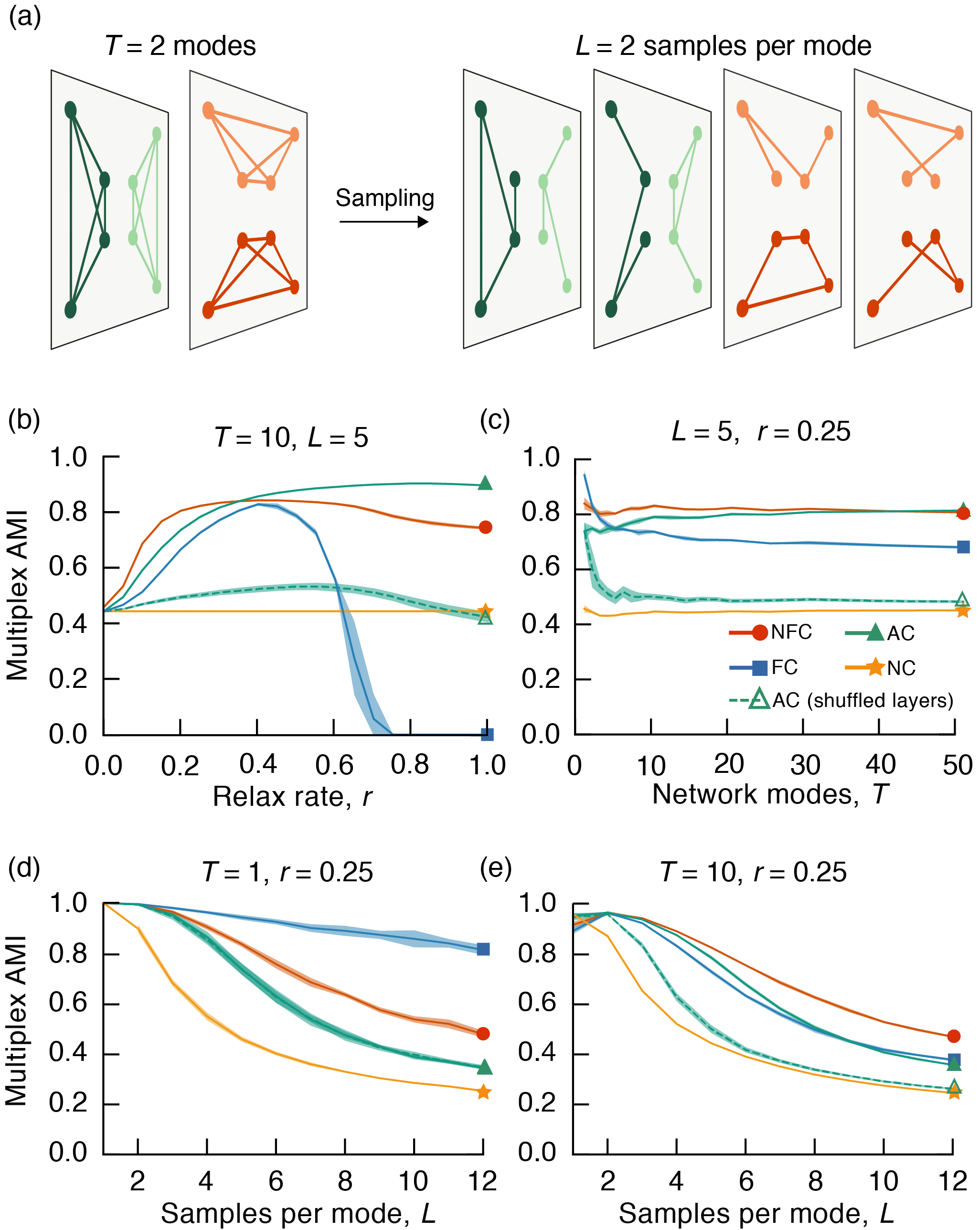}
    \caption{\label{fig:figures_network_modes_and_samples} \textbf{Neighborhood flow coupling captures intermittent overlapping communities in sparse multilayer networks}.
(a) Illustration of benchmark network for measuring performance in sparse networks with overlapping intermittent communities in $T$ network modes and $L$ sampled layers per mode.
(b) Performance of all coupling methods measured by the AMI between the recovered and true partitions as a function of the relax rate $r$. Per definition, no-coupling does not depend on the relax rate. (c) Performance test for $L=5$ layers per mode and increasing number of network modes $T$. (d-e) Performance test for fixed number of network modes $T=1$ and $T=10$, respectively, and increasing number of increasingly sparse sampled layers $L$ per network mode. For more than one network mode, neighborhood flow coupling stands out as the best coupling method.}
\end{figure}

\subsubsection*{Intermittent communities}
In networks with many layers, communities may persist over some period of time, then vanish and reemerge again by activating the same subset of nodes with similar within-group link structures. 
When the goal is to identify such intermittent communities, it is important to avoid spurious community merges. 
Therefore, we compare how different coupling schemes perform with respect to detecting intermittent communities in increasingly sparse multi-mode benchmark networks~\cite{de2015identifying}. 

First, we generate $T$ independent network layers, which we refer to as \textit{modes}, with the LFR benchmark model~\cite{lancichinetti2008benchmark}. To approximate the real networks that we study, each mode has $512$ nodes, average degree $8$, mixing coefficient $0.05$, and power-law community-size and degree distributions with exponent $3$ (see Appendix \ref{app:benchmark_networks} for more details).
From each mode we independently sample $L$ network layers that include links from their mode with probability $1/L$.
Each multilayer benchmark network thus comprises $T \times L$ layers, with $T$ independent sets of $L$ dependent layers, as schematically illustrated in Fig.~\ref{fig:figures_network_modes_and_samples}a. 
With increasing $L$ and sparser communities, the challenge is to detect the communities planted in each mode and distinguish between communities from different modes. 

To measure performance, we compute the adjusted mutual information~\cite{vinh2010information}, AMI, between the predicted and true state node labels. 
We first show that neighborhood flow coupling is less sensitive to variations in the relax rate (Fig.~\ref{fig:figures_network_modes_and_samples}b). 
The no-coupling method is independent of the relax rate and serves as a performance baseline. 
For adjacent coupling, the performance increases with the relax rate because this coupling takes advantage of the ordered layers and completes information in sparse layers. 
However, when shuffling the layers, this advantage vanishes and the performance drops significantly.  
Full coupling has a narrow performance optimum and the performance drops to zero around $r=0.7$ when the strong interlayer coupling causes Infomap to label the whole network as one community. 
Neighborhood flow coupling is more stable and performs best with a relax rate between $0.15$ and $0.7$ for this type of multilayer network. 
If not stated otherwise, we use $r=0.25$ for all analysis.

For all types of coupling, performance depends on the number of network modes.
On single-mode multilayer networks, full coupling scores the highest because uniform interlayer coupling maximally aggregates the dependent layers (Fig.~\ref{fig:figures_network_modes_and_samples}d). 
When the number of samples per mode increases, the networks become sparser and the probability of finding high-similarity neighborhood flows decreases. As a result, neighborhood flow coupling converges to no-coupling. 
However, neighborhood flow coupling handles many samples per mode and multiple modes better than any other coupling scheme. 
For example, in the ten-mode multilayer networks with overlapping communities, neighborhood flow coupling performs better than full coupling also for many samples per mode (Fig.~\ref{fig:figures_network_modes_and_samples}e).
In this case, both coupling schemes perform better for a few samples per mode than only one, because they force interlayer links between spuriously overlapping layers.
This behavior suggests that an adaptive relax rate based on the absolute similarity between layers may give even better results.
Nevertheless, the performance of neighborhood flow coupling remains stable for much higher numbers of network modes (Fig.~\ref{fig:figures_network_modes_and_samples}c).
While adjacent coupling performs on par with neighborhood flow coupling in this scenario, its performance rely on the layer order. When we shuffle the layers, adjacent coupling can no longer benefit from similarities between adjacent layers and perform as bad as no coupling. This result highlights that adjacent coupling cannot detect communities with temporal interruptions.
Overall, while there is room for further improvement, neighborhood flow coupling stands out as the best method for detecting intermittent communities.



\subsubsection*{Flow persistence} 
\begin{figure}
    \centering
    \includegraphics[width=\linewidth]{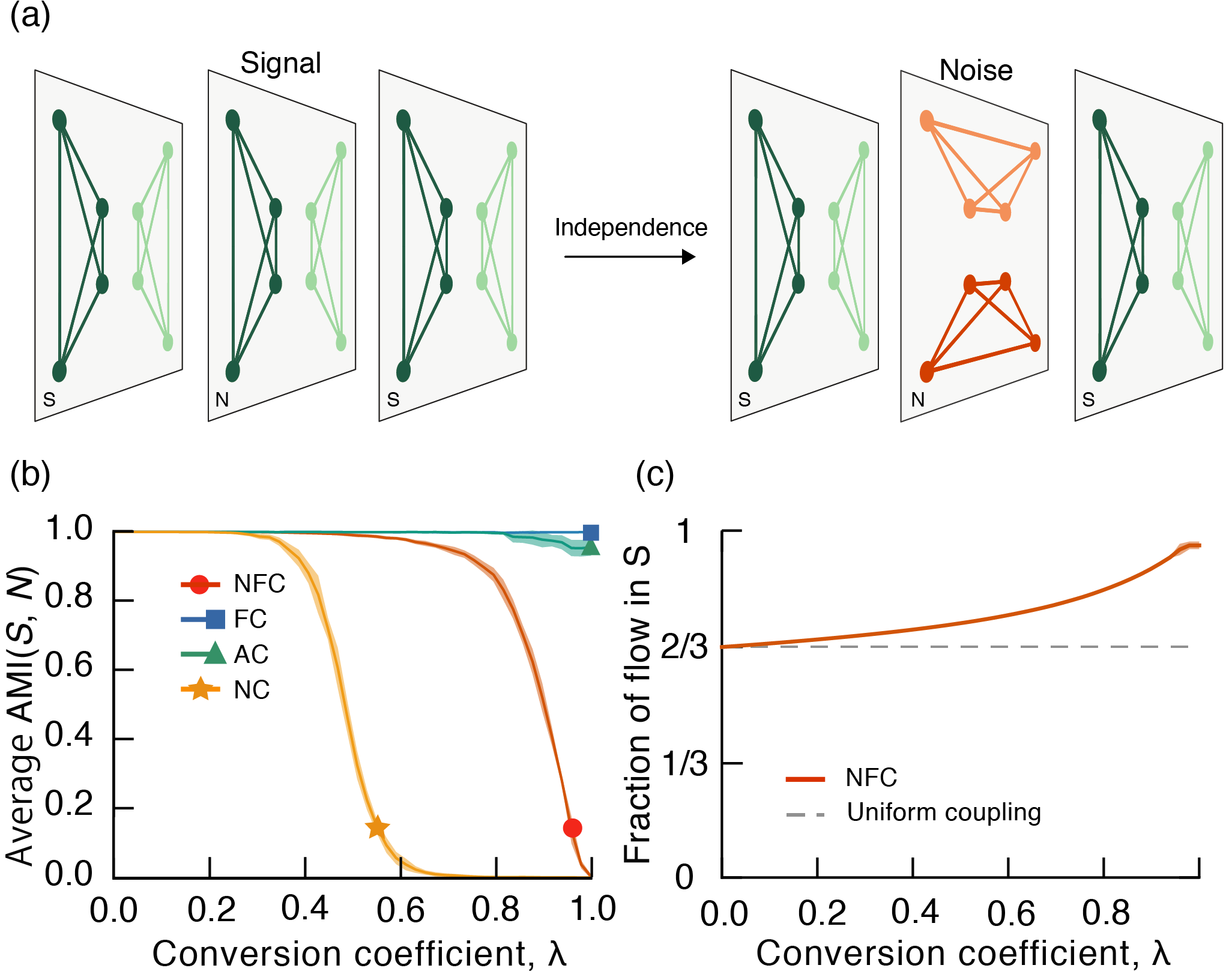}
    \caption{\label{fig:figures_information_leakage_and_flow_isolation} \textbf{Neighborhood flow coupling amplifies flow persistence in communities}. (a) Schematic benchmark network for testing network flow persistence in planted communities. Two identical signal layers sandwich a noise layer with varying overlap, with the signal layers set by conversion coefficient $\lambda$, from identical at $\lambda=0$ (left) to independent at $\lambda=1$ (right). (b) The average AMI between the identified partitions in the signal and noise layers decreases as they disentangle. (c) The fraction of total network flows captured by nodes in the signal layers increases as the signal and noise layers disentangle for neighborhood flow coupling, but is constant at two-thirds of the total flow for the uniform coupling methods. We use relax rate $r=1$ for neighborhood flow coupling to maximize interlayer flows and emphasize the effect.}
\end{figure}

We have developed the neighborhood flow coupling to constrain flows within structurally similar overlapping regions of a network.
To explore this feature, we use a multilayer benchmark network model consisting of two identical signal layers with known clusterings at both sides of a noise layer that, to a tunable degree, is more or less independent of the signal layers (Fig.~\ref{fig:figures_information_leakage_and_flow_isolation}a).
We generate layers with the same LFR benchmark model as in the previous section.
We introduce a tuning parameter $\lambda \in [0, 1]$ such that the noise layer contains $n_e(1 - \lambda)$ randomly selected edges from the signal network and likewise $n_e\lambda$ from another network generated independently following the same procedure. 
By tuning $\lambda$ from 0 to 1, we can gradually convert the noise layer from the signal network copy to an independent network~\cite{aldecoa2013exploring}. 
We can now test how well different coupling methods handle interference from the noise layer by measuring the decrease in average adjusted mutual information between the identified signal and noise layer partitions as we increase $\lambda$. To emphasize the effects, we use relax rate $r=1$.

Neighborhood flow coupling and no-coupling are robust to interference from irrelevant layers.
At some level of conversion, noise and signal layer communities should be considered independent of each other and the AMI between signal and noise layers should go to zero.
No-coupling gives independent labels to the noise layer after 60 percent conversion, and neighborhood flow coupling gives independent labels after 100 percent conversion.
Full and adjacent coupling suffer from interference with the noise layer even when it is fully converted and thus independent of the signal layers (Fig.~\ref{fig:figures_information_leakage_and_flow_isolation}b).
The strong coupling between signal and noise layers for these methods induces interlayer flows in spurious communities.
Obviously, the no-coupling method is immune to such interference, and therefore is unable to pick up actual interlayer coupling in intermittent communities (Fig.~\ref{fig:figures_network_modes_and_samples}). 
In contrast, neighborhood flow coupling is able to both avoid interference from irrelevant structures and pick up information from intermittent communities.

Neighborhood flow coupling can retain flows in intermittent communities. 
The proportion of flow inside the signal layers explains why neighborhood flow coupling outperforms full and adjacent coupling. 
In this three-layer example, for any uniform coupling scheme -- be it full, adjacent or no-coupling -- each layer carries one third of the total flow, independent of $\lambda$. 
Therefore, two-thirds of the total flow in the signal layers forms a baseline. 
For neighborhood flow coupling, however, this fraction increases as $\lambda$ approaches 1 and the signal and noise layers disentangle (Fig.~\ref{fig:figures_information_leakage_and_flow_isolation}c).
The adaptive coupling reinforces flows inside the two signal layers together and prevents flows from leaking to the noise layer. 
As a result, neighborhood flow coupling accentuates structures with long flow persistence times across layers and makes it possible to detect intermittent communities in multilayer networks.

\subsection*{Understanding real-world temporal contact networks} 
We now apply multiplex Infomap using neighborhood flow coupling, full coupling, adjacent coupling and no-coupling schemes to two empirical temporal contact networks. 
We represent each data set as a multilayer network and aggregate links over 10-minute intervals in each layer.
The first network represents contact events during working hours (approximately 8 a.m. to 6 p.m) between employees in a workplace environment over two weeks~\cite{genois2015data}.
In this network there are $n=92$ physical nodes, $e=2.91\cdot10^{3}$ intralayer links, $t=575$ non-empty layers, and the average intralayer node degree is $\left<k\right>=0.110$.
The second network arises from Bluetooth signal connections between personal smartphones of freshmen university students, also over two weeks~\cite{stopczynski2014measuring} ($n=636$, $e=1.27\cdot10^{5}$, $t=600$, $\left<k\right>=0.665$). 
In the university dataset, links are tracked during a special study period where each student attends the same course every day.
The students may meet anytime during the 24 hours of the day, but to simplify the comparison to the workplace network, we only consider links that occur during working hours (8 a.m.--4 p.m.).
Thus, both networks are cropped to this time-frame so $t=480$.
We start by analyzing the interlayer link structure that neighborhood flow coupling produces.
In particular, we are interested in understanding the sparsity of the representations that the method creates, compared to other methods.
We then evaluate the performance of Infomap resulting from each coupling scheme by measuring overlap, size, and self-similarity over the time of communities that each method finds.
There is no ground truth to measure performance against, so we focus our analysis on showing that neighborhood flow coupling strikes a balance between allowing information to flow between all layers -- the strength of full coupling -- and not mixing unrelated contexts -- the strength of no-coupling.
Finally, we explore each network by visualizing the neighborhood flow coupling community detection solution.

\subsubsection*{Neighborhood flow coupling finds communities that are highly self-similar}
\begin{figure}
    \centering
    \includegraphics[width=\linewidth]{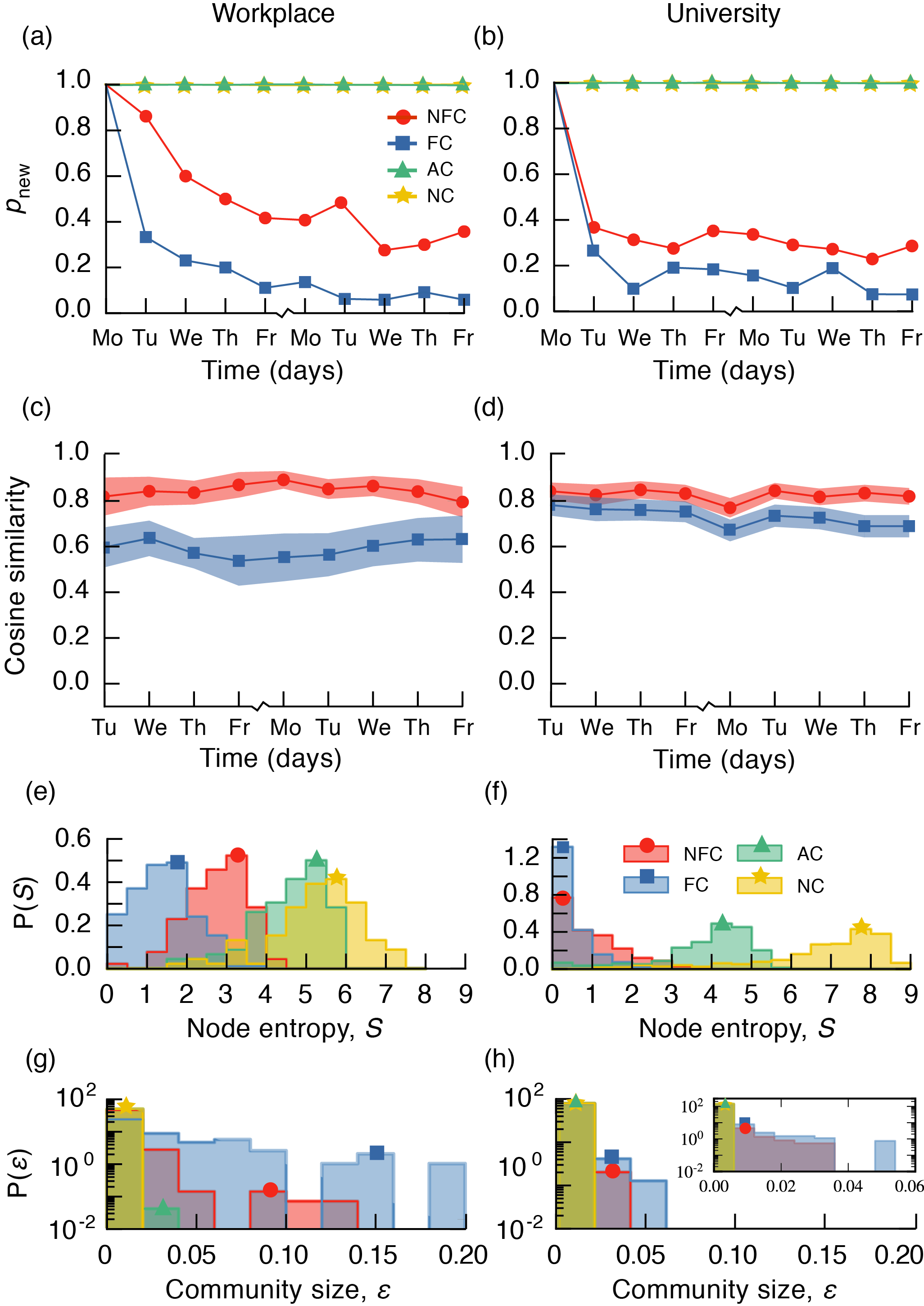}
    \caption{\label{fig:community_reoccurance}\textbf{Properties of communities in real-world networks}. Each column corresponds to a separate dataset. (a-b) The number of new communities on each day, $p_{\mathrm{new}}$, in the solutions for each coupling scheme. (c-d) Average community self-similarity over time. (e-f) Distribution of node community assignment entropy, $S$, as a measure of community overlap. (g-h) Community size distributions.}
\end{figure}

We know from the literature that these networks contain intermittent communities~\cite{sekara2016fundamental, stehle2011high}.
Therefore these networks are useful to better understand each different coupling method's ability to capture intermittent community structure.
Due to the frequent daily re-emerging of communities, a method that couples temporally distant layers should cause the rate of new communities discovered on each day, $p_{\mathrm{new}}$, to decline over time.
This is indeed the case for full coupling and neighborhood flow coupling (Fig. \ref{fig:community_reoccurance}a-b).
Full coupling drives $p_{\mathrm{new}}$ close to zero, which is unrealistic as we should expect some degree of exploration to take place.
The reason for this behavior is likely the fact that new communities are merged with previous, slightly overlapping communities.
For neighborhood flow coupling, intermittent communities are appropriately recognized each day, while a significant fraction of new configurations is given new labels.

Knowing that communities are indeed successfully rediscovered each day, we now seek to understand how self-similar intermittent communities are between days of (re)discovery.
A good detection algorithm should partition the network such that each reappearance of a community is highly self-similar to its other appearances.
We measure the similarity between each temporal community to itself on the most recent previous day as the cosine similarity between the unnormalized 24-hour aggregate distributions of member nodes, and plot the similarity distribution over time as their means inside the 95\% confidence intervals.
It is only relevant to measure self-similarity for full and neighborhood flow coupling, since only those two methods are able to capture reoccurring communities.
In both networks, neighborhood flow coupling results in, on average, higher community self-similarity than full coupling does (Fig. \ref{fig:community_reoccurance}c-d).
This difference is more pronounced in the university network because the structures are larger.
In the case of full coupling, large communities are frequently split into smaller ones that are rarely detected.

\subsubsection*{Full-coupling solutions tend to merge overlapping communities}
We measure the distribution of node entropy, $P(S)$, and the distribution of community size, $P(\epsilon)$, in each network. 
We compute the node entropy as $S=\sum_ic_i\log{c_i}$ where $c_i$ is the distribution over time spent in community $i$ for a given node.
Intuitively, if the average node entropy is high, nodes are detected as frequently being in different communities, meaning that communities must overlap on many nodes.
Full coupling results in low node entropy and large communities (see Fig. \ref{fig:community_reoccurance}c-e). 
In conjunction with our previous observation that full coupling leads to unrealistically low values of $p_{\mathrm{new}}$, this is a strong indication that it causes Infomap to merge communities that overlap in different layers.

For both networks, the $p_{\mathrm{new}}$ curve for neighborhood flow coupling is similar to full coupling but with more new communities emerging each day. 
In the workplace network, we note that there are almost as many new communities discovered on the second day as there are on the first. 
We can explain this with the observation that the workplace dataset contains groups that are scheduled to meet every other day and, as such, we should expect some of those to start on the second day. 
While neighborhood flow coupling captures this nuance, full coupling does not. 
In the university network, neighborhood flow coupling identifies fewer and fewer communities as the week progresses, with the exception of Fridays, where relatively more new communities form.
This nuance is not captured by full coupling.
These results further support the concept that full coupling results in mergers of overlapping communities due to interlayer links that connect them via the nodes they overlap on.

\begin{figure}
    \centering
    \includegraphics[width=\linewidth]{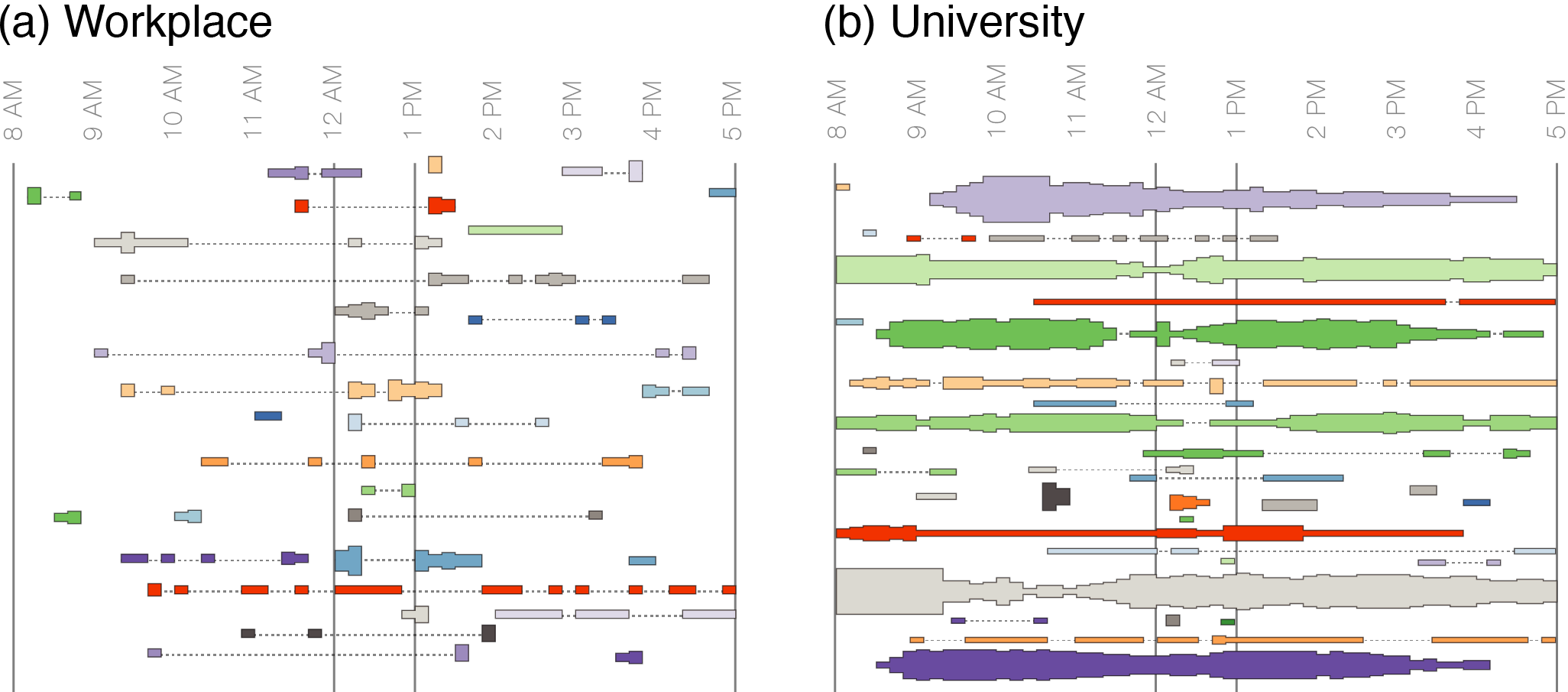}
    \caption{\label{fig:sausages_NFC} \textbf{Temporal communities detected by Infomap with neighborhood flow coupling}. Each horizontal track represents a community and its varying height represents the number of active nodes over time. (a) Partition of the workplace network. Height to scale with (b). (b) Partition of the university network. At its tallest point, the largest community (top purple, 10 am) has 22 active members.}
\end{figure}

\subsubsection*{Visualization of temporal communities}
We visualize the temporal expansion and contraction of communities found by Infomap with neighborhood flow layer coupling in each network, as horizontal "strips" of varying height~\cite{jensen2016}. 
Figure~\ref{fig:sausages_NFC} displays a subset of the communities discovered in each network (vertical position is arbitrary).
There are clear differences between the community structure of the two networks.
The university network gives rise to large structures that persist over long periods of time, while the workplace communities are significantly more intermittent, lasting on the order of tens of minutes.
Community sizes agree with our insight from Figure~\ref{fig:community_reoccurance}g-h. 
In the university network, some are large, corresponding to students attending lectures, some are mid-sized, corresponding to work-groups and small lectures, and some are small, corresponding to 2--4 person gatherings.
In the workplace network, communities mostly consist of a few people and occasionally are larger around lunch, but never in a scale similar to the university, as we should expect.
We provide an interactive version of Figure~\ref{fig:sausages_NFC} at \url{http://ulfaslak.com/research/temporal_communities/}, which offers further intuition about these networks and the effects of neighborhood flow coupling to the observed structure.
With these levels of intermittent communities -- here observed in particular for the workplace network but also strongly present in the university network at a daily rate -- it is clear that neighborhood flow coupling is a good choice for estimating layer interdependency. 

\section*{Conclusion}
Our experiments suggest that connecting state nodes across layers in multilayer networks based on the similarity between their network neighborhood flows has multiple benefits over uniform entire-layer coupling approaches. 
For example, in series of timestamped face-to-face interaction events represented as multilayer networks, neighborhood flow coupling captures natural constraints on information flows such that flows move freely only within and between similar communities across layers. 
As a result, Infomap is able to identify intermittent communities with long flow persistence times and recognize spuriously overlapping communities as separate entities.
In contrast, existing uniform entire-layer approaches either fail to capture whole communities that are intermittent across temporal layers or collapse spuriously overlapping communities into single communities.
Furthermore, we demonstrate that neighborhood flow coupling results in multilayer network representations that are orders of magnitudes sparser in typical real-world networks with corresponding computational gains.
This computational gain allows us to analyze and identify intermittent communities in temporal networks over longer times or higher resolution. 
Consequently, neighborhood flow coupling opens new avenues for temporal network analysis. 

\begin{acknowledgements}
We thank Chris Bl{\"o}cker and Ludvig Bohlin for comments that improved our manuscript.
M.R. was supported by the Swedish Research Council, grant 2016-00796. SL was funded by the Danish Council for Independent Research, grant number 4184-00556a, and the Villum Foundation `High resolution networks'. University contact data (from the Copenhagen Networks Study~\cite{stopczynski2014measuring}) was collected with the permission of the Danish Data Protection Agency (Journal number 2012-41-0664).
\end{acknowledgements}

\bibliography{bibliography}

\appendix

\section{Interlayer sparsity}
To couple state nodes, neighborhood flow coupling requires that the network structure around state nodes be similar. 
This similarity is not required by full coupling, which couples state nodes of each physical node regardless.
In temporal networks, there are often many state node pairs that have no structural similarity, because they can participate in non-overlapping communities at different times.
With neighborhood flow coupling, this results in network representations with sparse interlayer link structure compared to full coupling, where the interlayer network is always dense.
In this appendix, we investigate the degree to which neighborhood flow coupling reduces the size of the network, and thus the memory footprint.
We measure the density reduction in relation to the full-coupling density, which is always one, and compare to the adjacent coupling density.
Furthermore, we analyze how density varies with layer interdependence for neighborhood flow coupling, as we reason that this must be an important factor.

First, we consider sparsity in synthetic networks with independent layers. We define interlayer density, $S$, as the ratio of realized to possible interlayer links.
Per definition, it is always the case that $S_{FC}=1$ and $S_{NC}=0$.
If layers are independent, we can derive $S_{AC}=2/t$ by dividing the expected number of links from adjacent coupling with the expected number of links from full coupling.
For $t=600$ (corresponding to two weeks of working hours in 10-minute time-bins), $S_{AC}=3.3\cdot10^{-3}$. 
$S_{NFC}$ can be approximated as the probability that two state nodes have at least one link in common:
\begin{align}
S_{NFC}(\left<k\right>, n) = 1 - P\left(0, \frac{\left<k\right>^2}{n}\right)  = 1 - e^{-\left<k\right>^2/n},
\label{eq:s_nfc}
\end{align}
where $P(0, \theta)$ is the function value in 0 of a Poisson distribution with average $\theta$ equal to the expected number of shared links between two state nodes in independent layers $\left<k\right>^2/n$.
For a network with similar statistics to the university network ($n=636$, $\left<k\right>=0.665$), Eq. \eqref{eq:s_nfc} gives $S_{NFC}=7.0 \cdot 10^{-4}$.
In real temporal networks, however, we observe that the interlayer link structure is more dense because there is significant dependence between layers.
The estimations presented here therefore only serve as random network baselines that we can compare with.
For the university network, where the possible number of interlayer links is $7.11\cdot10^{7}$, we observe $S_{NFC}=0.680$ and $S_{AC}=0.005$, and for the workplace network, where the possible number of interlayer links is $4.33\cdot10^{5}$, $S_{NFC}=0.299$ and $S_{AC}=0.012$.
The large increase in $S$ that we observe for the empirical networks reveals that neighborhood flow coupling is very sensitive to interdependence between layers.

\begin{figure}
    \centering
    \includegraphics[width=\linewidth]{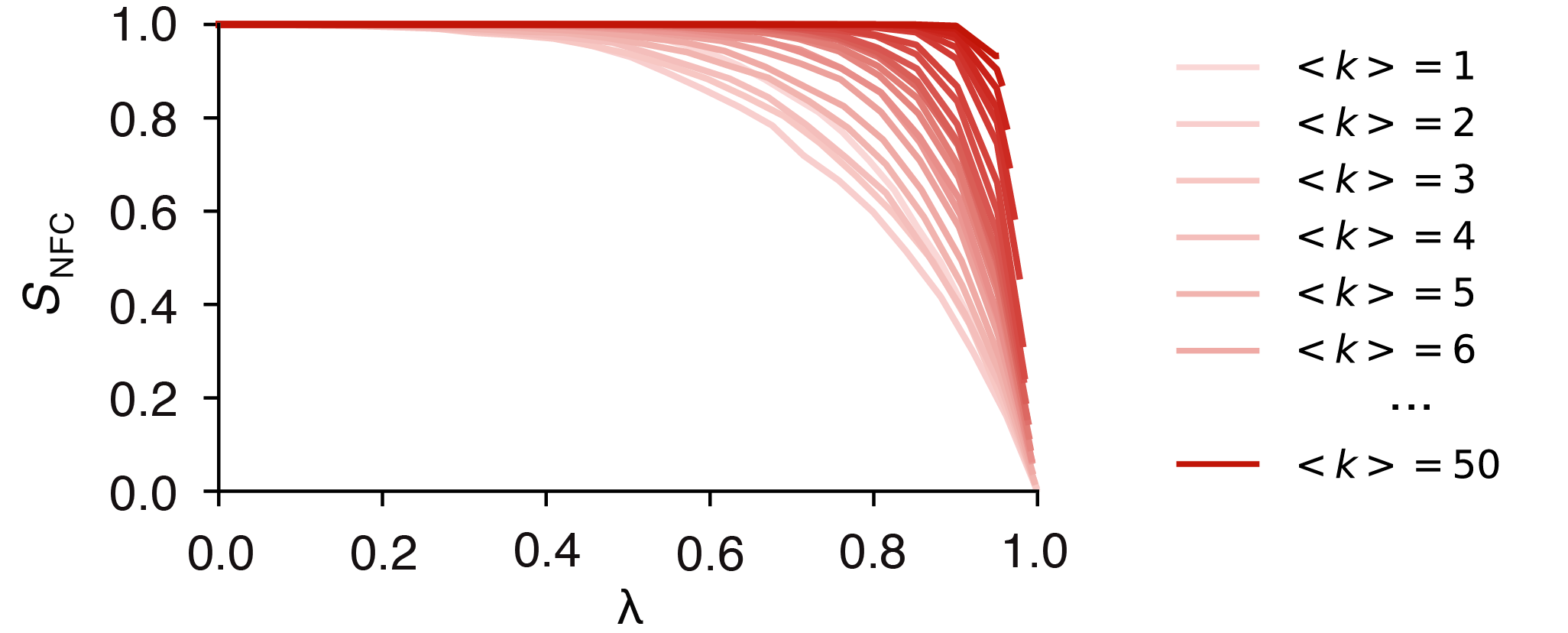}
    \caption{\label{fig:interlayer_sparsity} \textbf{Interlayer sparsity depends on layer interdependence.} The density of interlayer links $S_{NFC}$ created by neighborhood flow coupling decreases as the layers in a multilayer network become more independent (increasing $\lambda$). The rate at which density decreases depends on the average degree of state nodes $\left<k\right>$. Sparse intralayer link structure, corresponding to low $\left<k\right>$, leads to sparser interlayer link structure.}
\end{figure}

We now test how sensitive interlayer sparsity resulting from neighborhood flow coupling is to layer interdependence, using a simple experiment similar to the approach taken in the \textit{Flow persistence} section above. 
We create an Erd\H{o}s-R\'enyi graph with $n=1000$ and variable $\left<k\right>$.
We create a two-layer network where both layers are copies of this network, such that the layer independence $\lambda$, which we measure as the average Jensen-Shannon divergence across all pairs of state nodes, is zero.
We then gradually convert the second layer to an independent network, generated by the same process, using edge swaps, while measuring $S_{NFC}$ versus $\lambda$.
When the second layer is fully converted, the two layers are maximally independent and $\lambda=1$.
The experiment shows, first of all, that the relationship between $S_{NFC}$ and $\lambda$ is non-linear. 
Secondly, we observe that a sparse intralayer structure (low $\left<k\right>$) leads to a sparser interlayer link structure, increasingly so when layers are independent (Fig. \ref{fig:interlayer_sparsity}).

Thus neighborhood flow coupling offers significant gains in memory efficiency relative to full coupling, particularly in sparse multilayer networks.

\section{Robustness to relax rate}
In absence of an adaptive relax rate, the problem at hand should decide what relax rate $r$ to use. In general, for full coupling, $r$ must be large enough to facilitate flow between layers yet small enough to contain information inside the layer communities.
For neighborhood flow coupling, this heuristic does not apply, because interlayer links are established only between structurally similar regions of the network.
At the same time, $r$ still controls the amount of interlayer flow in the network.
If $r=0$, information cannot flow between layers, and if $r=1$, important layer information may be diluted.

The optimal relax rate $r$ should allow Infomap to discover communities that repeat in different layers.
To test this criterion, we perform a simple experiment that starts with a multilayer network, selects a random layer and appends a copy of it to the end of the network.
For a range of $r$ values, we then measure the proportion of nodes in the copied layer to which Infomap assigns the same label as in the original layer.
We perform this test on the university and the workplace networks for  neighborhood flow and full coupling, and find that both coupling schemes give perfect labeling of all copied nodes for all values of $r$ except $r=0$.
While this result does not reveal a performance optimum for $r$, it shows that the map equation can effectively capture layer interdependences.

\begin{figure}
    \centering
    \includegraphics[width=\linewidth]{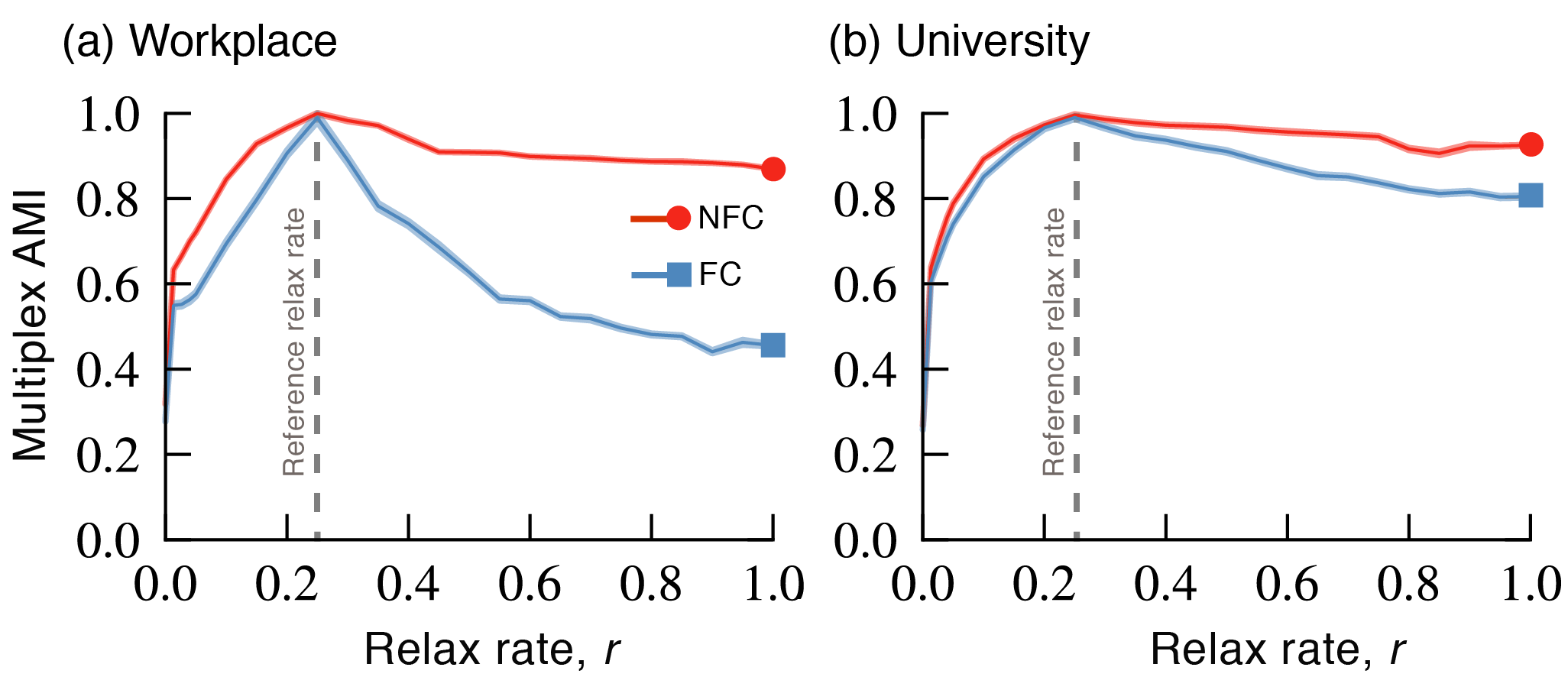}
    \caption{\label{fig:figures_r_robustness}\textbf{Neighborhood flow coupling is highly robust to variations in $r$.} The plots illustrate the similarity of a community detection solution using $r=0.25$ with solutions obtained from different relax rates. The high values for neighborhood flow coupling in both networks (a) and (b) demonstrate its high robustness to $r$ variability compared to full coupling.}
\end{figure}

The results should not be sensitive to the exact choice of the relax rate.
We demonstrate the robustness by clustering a network for a range of relax rates and comparing each solution to the solution for $r=0.25$, with the multiplex AMI as a performance measure.
If robustness is high, all solutions should have a high AMI with this reference solution.
Performing this test for both networks, we find that neighborhood flow coupling solutions are significantly more robust to varying $r$ than full-coupling solutions.
Neighborhood flow coupling is particularly robust in the domain $r>0.25$.
The similarity decays faster when $r<0.25$ and goes to zero for $r=0$, which demonstrates that, while robust to $r$, Infomap with neighborhood flow coupling allows for detecting smaller communities. In summary, a broad spectrum of relax rates gives similar solutions for Infomap with neighborhood flow coupling (Fig. \ref{fig:figures_r_robustness}). 

\section{Benchmark networks}
\label{app:benchmark_networks}

\begin{figure}
	\centering
    \includegraphics[width=\linewidth]{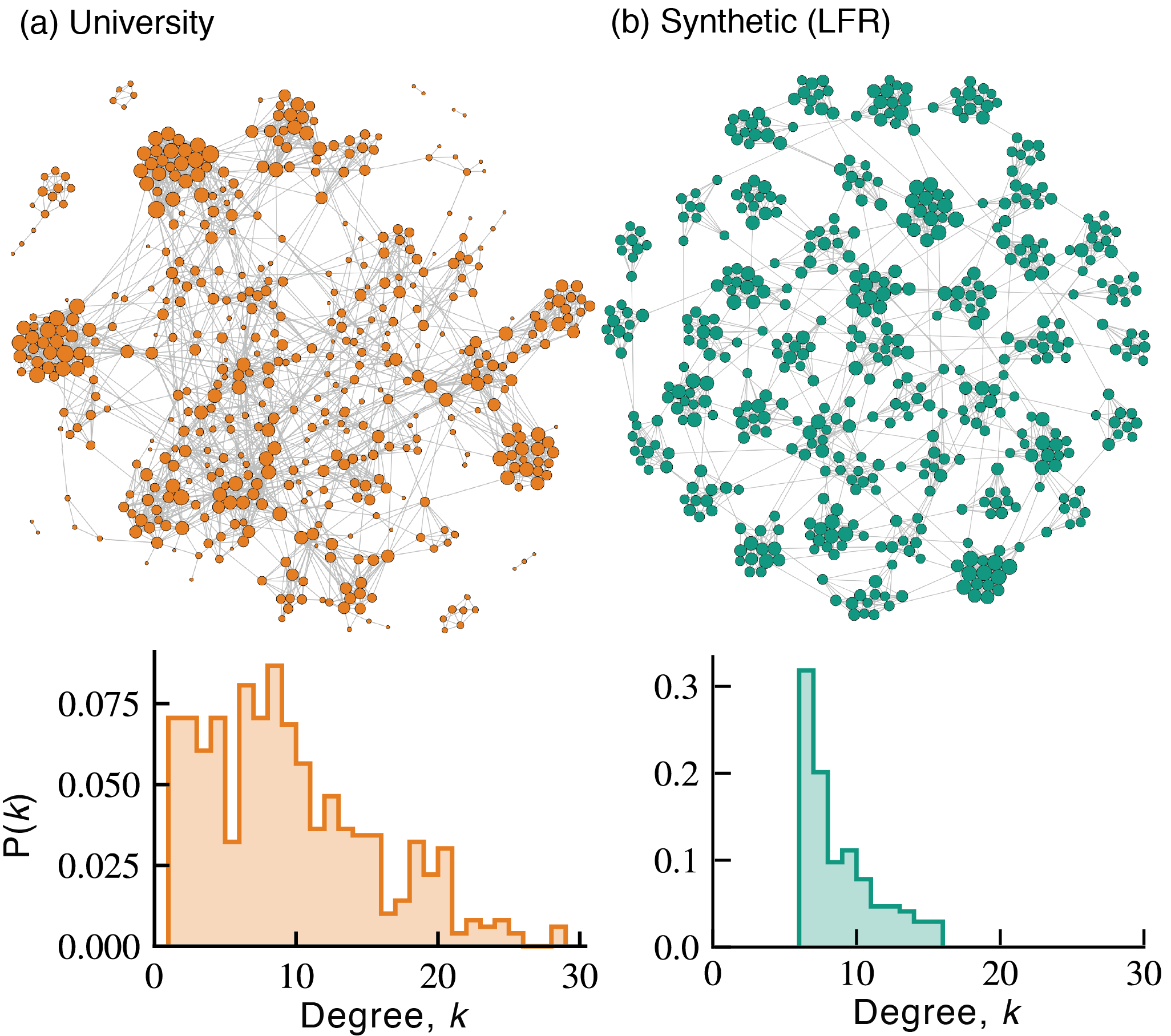}
    \caption{\label{fig:figures_benchmark_and_real_networks}\textbf{Real and synthetic networks}. (a) Simple network of aggregated links between 8 a.m.\ and 4 p.m.\ in the university dataset. The number of nodes is 496 and the mean degree is 9.0. (b) Synthetic LFR network realization with 512 nodes, average degree 8.4 and fitted degree distribution power law exponent 3.8.}
\end{figure}

For inferring results from benchmark networks to real networks, the benchmark networks should resemble the real networks. However, real networks come in a great variety and benchmark networks cannot accurately mimic all of them. To find meaningful model parameters for the LFR benchmark networks~\cite{lancichinetti2008benchmark}, we consider individual workdays of the university network as aggregated simple graphs (interactions between 8 a.m.~and 4 p.m.) and observe that the number of nodes typically lies between 400 and 500 and that the mean degree is in the range 6--12.
Figure \ref{fig:figures_benchmark_and_real_networks}a shows an example of one such real network and its degree distribution.
We generate synthetic networks with the LFR implementation made available online~\cite{lfrcode}, with input parameters: $N=512$ (number of nodes), $k=8$ (average degree), \texttt{maxk} $=16$ (maximum degree), $\mu=0.05$ (mixing), \texttt{t1} $=$ \texttt{t2} $=3$ (degree and community-size power-law distribution exponent) and \texttt{maxc} $=24$ (maximum community size).
While there is no guarantee that the statistics of individual resulting networks fully respect the input parameters, we observe that realized degree and power law exponents deviate only marginally (standard deviations 0.1 and 0.8 respectively).
Figure \ref{fig:figures_benchmark_and_real_networks}b shows an example of a synthetic network that results from these parameters.

\end{document}